\documentclass[manuscript]{emulateapj} 
\usepackage{natbib}

\usepackage{color}
\usepackage{graphicx}
\usepackage{soul}

\newcommand{\be}{\begin{equation}}
\newcommand{\ee}{\end{equation}}
\newcommand{\bea}{\begin{eqnarray}}
\newcommand{\eea}{\end{eqnarray}}

\newcommand{\lum}{\mathcal{L}}
\newcommand{\om}{{\Omega_{\rm m}}}
\newcommand{\ho}{{H_0}}

\newcommand{\Planck}{{\sl Planck\/}} % use \Planck\ in the text

\begin{document}

\title{Evidence of cross-correlation between the CMB lensing and the  $\gamma$-ray sky}

\author{Nicolao Fornengo}
\affiliation{Dipartimento di Fisica, Universit\`a di Torino, I-10125 Torino, Italy}
\affiliation{Istituto Nazionale di Fisica Nucleare, Sezione di Torino I-10125 Torino, Italy}

\author{Laurence Perotto}
\affiliation{LPSC, Universit\'e Grenoble-Alpes, CNRS/IN2P3, 53, rue des Martyrs 38026 Grenoble Cedex, France}

\author{Marco Regis}
\email{regis@to.infn.it}
\affiliation{Dipartimento di Fisica, Universit\`a di Torino, I-10125 Torino, Italy}
\affiliation{Istituto Nazionale di Fisica Nucleare, Sezione di Torino I-10125 Torino, Italy}

\author{Stefano Camera}
\affiliation{Jodrell Bank Centre for Astrophysics, The University of Manchester, Manchester M13 9PL, UK}
\affiliation{CENTRA, Instituto Superior T\'ecnico, Universidade de Lisboa, Lisboa, Portugal}

\date{\today}

\begin{abstract}
We report the measurement of the angular power spectrum of cross-correlation between the 
unresolved component of the {\sl Fermi-LAT} $\gamma$-ray sky-maps and the CMB lensing potential map reconstructed by the \Planck\ satellite.
The matter distribution in the Universe determines the bending of light coming from the last scattering surface.
At the same time, the matter density drives the growth history of
astrophysical objects, including their capability at generating
non-thermal phenomena, which in turn give rise to  $\gamma$-ray emissions.
The \Planck\ lensing map provides information on the integrated distribution of matter, while the integrated history of  $\gamma$-ray emitters is imprinted in the {\sl Fermi-LAT} sky maps.
We report here the first evidence of their correlation.
We find that the multipole dependence of the cross-correlation
measurement is in agreement with current models of the  $\gamma$-ray luminosity function for AGN and star forming galaxies,  with a statistical evidence of 3.0$\sigma$.
Moreover, its amplitude can in general be matched only assuming that these extra-galactic emitters are also the bulk contribution of the measured isotopic $\gamma$-ray background (IGRB) intensity.
This leaves little room for a big contribution from galactic sources to the IGRB measured by {\sl Fermi-LAT}, 
pointing toward a direct evidence of the extragalactic origin of the IGRB.
\end{abstract}

\pacs{98.80.-k, 98.80.Es, 98.70.Rz, 98.62.Sb}

\maketitle

%%%%%%%%%%%%%%%%%%%%%%%%%%%%%%%%%%%%%%%%%%%%%%%%%%%%%%%%%%%%%%%%%%%%%%%%%%%%%
\section{Introduction}
\label{sec:intro}

The weak gravitational lensing by large scale structures imprints the integrated dark matter distribution onto the cosmic microwave
background (CMB) anisotropies. It results in a remapping of the CMB observables, which depends on the line-of-sight integral of the gravitational potential, with a broad kernel peaking at 
a redshift $z \sim 2$, and which is referred to as the \emph{lensing potential} \citep{Blanchard:1987} (see also \citep{Lewis:2006fu} for a review). This process perturbates the statistical properties of the CMB observables, which are primarily very close to Gaussian fields. This non-Gaussian signature can be exploited to extract the lensing potential from the CMB maps~\citep{Okamoto:2003zw}. Using such a technique, the \Planck\ Collaboration obtained nearly all-sky maps of the lensing potential reconstructed from an undusted CMB  temperature map~\citep{Planck2013lensing} and from both foregorund-cleaned CMB  temperature and polarisation maps~\citep{Planck2015lensing}. They provide us with an estimate of the matter distribution, mainly sensitive to halos located at $1\lesssim z\lesssim3$.

On the other hand, the accretion of baryonic matter in halos also creates active astrophysical objects. They can host violent phenomena, such as, e.g., supernova explosions and relativistic outflows, which are able to accelerate particles to high-energies.
Particles with GeV-TeV energy interacting with the ambient medium emit
$\gamma$-ray radiation, mostly by means of production and decay of
neutral pions, inverse Compton scattering, and non-thermal
bremsstrahlung. In addition, the same dark matter which forms the halos could produce  $\gamma$-rays, through its self-annihilation or decay.
In the past few decades, the all-sky diffuse $\gamma$-ray emission has been measured, but its origin and composition remain key open questions in high-energy astrophysics.
The featureless energy spectrum of the isotropic $\gamma$-ray background (IGRB)~\citep{Abdo:2010nz,Fermi:2014} and its flat angular power spectrum
(APS)~\citep{Ackermann:2012uf} make the IGRB identification a complex task. 
The cross-correlation of the IGRB with large scale structure tracers is a very valuable technique for understanding its composition~\citep{Xia:2011,Camera:2012cj,Fornengo:2013rga,Ando:2013xwa,Shirasaki:2014noa,Ando:2014aoa}.

In this work, we first show that the lensing potential map estimated by \Planck\ 
and the $\gamma$-ray sky observed by {\sl Fermi-LAT} do correlate, by reporting a measurement of their cross-correlation APS.
This stems from their common origin associated to extragalactic structures, and we discuss the extragalactic $\gamma$-ray background (EGB) properties which can explain the measurement.

The adopted cosmological model throughout this paper is the six-parameter $\Lambda$CDM \Planck\ best fitting model reported in \citep{Ade:2013zuv}.

%%%%%%%%%%%%%%%%%%%%%%%%%%%%%%%%%%%%%%%%%%%%%%%%%%%%%%%%%%%%%%%%%%%%%%%%%%%%%
\section{Data and Analysis}
\label{sec:data}

We use the $\gamma$-ray measurements obtained by the {\sl Fermi-LAT} in its first 68 months of operations,
from early August 2008 to late April 2014. We have processed the data with 
the {\sc Fermi Science Tools} version \texttt{v9r32p5}, using the Pass7-reprocessed  instrument
response functions for the {\sc clean} event class (P7REP\underline{\hspace{0.5em}}CLEAN\underline{\hspace{0.5em}}V15) for both {\sc front} and {\sc back} conversion types of events, which have been taken together. 
We have selected photon counts from 700 MeV to 300 GeV, subdivided into 70 energy bins (uniform in log scale) and mapped with a pixel size of 0.125$^\circ$ (suitable for subsequent {\sc HEALPIX} projection with $N_{\rm side} = 512$).  The {\sl Fermi-LAT} exposure maps have been derived on the same energy grid and resolution,
and we adopted a step size $\cos\theta=0.025$, in order to have sufficiently refined exposures.
From the count and exposure map cubes, we have finally derived the full-sky flux maps. For the cross-correlation analysis, we have grouped the energy sections in 6 bins (with boundaries at:
0.7, 0.99, 2.0, 5.1, 10.2, 48.7, 300 GeV).

The maps are contaminated by the galactic foreground: since we are interested in the extragalactic signal only, the maps have been cleaned by subtracting the {\sl Fermi-LAT} galactic model \texttt{gll\_iem\_v05}, which can be obtained from the {\sl Fermi-LAT} website\footnote{http://fermi.gsfc.nasa.gov/ssc/data/access/lat/ BackgroundModels.html}.

We account for Fermi-LAT PSF attenuation (which can be relevant at the angular scales of interest) by correcting the measured APS through a beam window function built as described in \citep{Ackermann:2012uf}.

As part of its public data releases of 2015 (2013), the \Planck\ Collaboration provided a CMB lensing convergence (potential) map that was employed in our analysis. 
We use the convergence harmonic coefficients $\kappa_{\ell m} = \ell(\ell+1)/2 \, \hat{\phi}_{\ell m}$ instead of the potential $\hat\phi$ to reduce the steepness of the APS as a function of multipoles.
The methodology followed to derive a convergence map from the 2013 potential map is described in \citep{Planck2013lensing}. 

We then mask regions contaminated by galactic foreground and
extragalactic sources. For the lensing maps, we adopt the publicly released masks, that both preverse about 70\% of the sky. We stress that these masks largely account for the galactic dust emission and the carbon-monoxide lines (that may correlate to $\gamma$-ray foreground). In order to mitigate against multipole mixing, we further use an apodisation over 5 degrees for the Planck 2013 analysis. 
For the $\gamma$-ray maps we prepare two masks combining the
\Planck\ 2013 lensing mask, a cut for galactic latitudes $|b|<25^\circ$
and excluding a $1^\circ$ angular radius around each source in the
2-year {\sl Fermi-LAT} catalog (2FGL)~\citep{Nolan:2012} and the 4-year {\sl Fermi-LAT} catalog (3FGL)~\citep{3FGL}, respectively. The 2FGL and 3FGL masks are
apodized over $3^\circ$ and $2^\circ$, respectively (the first choice being meant to provide us with a more conservative test), and the resulting effective sky fraction available is about 24\% (2FGL) and 23\% (3FGL).
We explored different apodizations and sets of galactic masks (including larger galactic cuts and an additonal mask for the region of the so-called ``Fermi Bubbles''~\citep{Su:2010}), finding consistent results.

The cross-correlation APS between the \Planck\ lensing map and the {\sl Fermi-LAT} $\gamma$-ray map is estimated using a pseudo-$C_\ell$ approach~\citep{Hivon:2001jp}.
To this aim, we make use of the publicly available tool {\sffamily PolSpice}~\citep{Chon:2003gx,Szapudi:2001}.
Although the {\sffamily PolSpice} algorithm properly deconvolves the signal APS from mask effects, it is known not to be a minimum variance algorithm~\citep{Efstathiou:2003dj}.
Thus the associated covariance matrix is likely to be an overestimation of the actual uncertainty, and the significances reported throughout the paper can in turn be considered as conservative.

We band-pass filter the cross-correlation APS in the multipole range $40<\ell<400$ in order to reduce possible contamination from systematic effects. This multipole range was defined in \citep{Planck2013lensing} as a confidence interval retaining $90\%$ of the lensing information, with multipoles $\ell<40$ requiring large \emph{mean-field} bias corrections. Similarly, multipoles above few hundreds are hardly accessible with the {\sl Fermi-LAT} sensitivity and angular resolution, and low multipoles correspond to the scales where the foreground cleaning has the largest impact~\citep{Ackermann:2012uf}.   
Since the expected signal is predicted to scale as $1/\ell$ (see next Section), the analysis is performed in terms of $\ell\,C_\ell^{(\gamma\kappa)}$.
In order to further mitigate mask mode-mixing, we average the APS in six linear bins of width $\Delta\ell=60$.

First, we measure the cross-correlation APS between the CMB lensing and a single $\gamma$-ray map derived from the integrated counts at $E>1$ GeV.
A hint of a signal in the low-$\ell$ range (with a peak at $\ell\lesssim 150-160$) is present, while the larger-$\ell$ bins are compatible with no deviation from a null signal.
We estimate the global significance of this low-$\ell$ peak evaluating the ratio of the measured APS over its error,
$\langle \ell C_\ell^{\gamma_i\kappa}\rangle/\delta\langle \ell C_\ell^{\gamma_i\kappa}\rangle$, in three multipole-bins: $40\leq\ell<160$, $160\leq\ell<280$, $280\leq\ell<400$.
The errors are the diagonal elements of the covariance matrix obtained from {\sffamily PolSpice}. We found the off-diagonal terms of the binned covariance matrix to be negligible.
Considering the Planck 2015 map with 3FGL mask (which is our reference analysis), the significances in the three multipole bins amount to:
$1.7\sigma$, $0.0\sigma$ and $0.2\sigma$, respectively.
The significance of the first bin for the four analyses arising from the combination of the Planck maps (2013 and 2015 releases) and the $\gamma$-ray point-source masks (2FGL and 3FGL) is reported in the first line of Table~\ref{tab:sigmas}.

\begin{figure}[t]
\vspace{-3cm}
\includegraphics[width=0.45\textwidth]{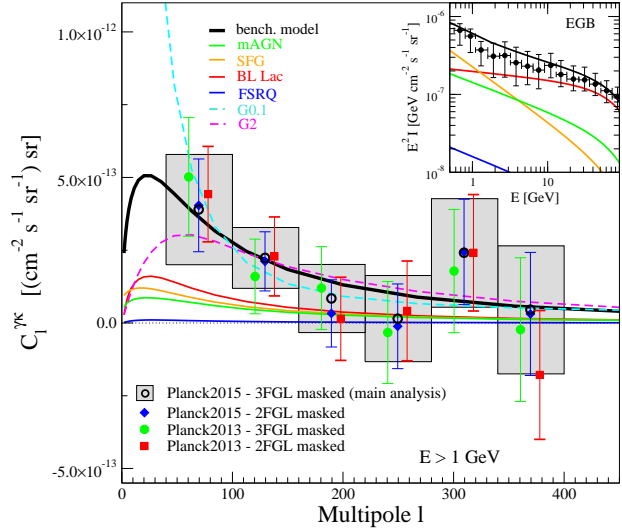}
\caption{
Cross-correlation APS $C_\ell^{(\gamma\kappa)}$ as a function of the multipole $\ell$, for $\gamma$-ray energies $E>1$ GeV.
The measurements are averaged (linearly in terms of $\ell\,C_\ell^{(\gamma\kappa)}$) in multipole bins of $\Delta \ell = 60$, starting at $\ell=40$.
Points report the minimum-variance combination of the measurement in individual energy bins (assuming a spectrum $\propto E^{-2.4}$), as described in Eq.~\ref{eq:mv}.
Four different analyses are shown. They arise from the combination of two lensing maps (from Planck 2013 and 2015 releases) and two $\gamma$-ray point-source masks (2FGL and 3FGL).
The benchmark theoretical model, shown in black, is the sum of the contributions from BL Lac (red), FSRQ (blue), mAGN (green), and SFG (orange), multiplied by $ A^{\gamma\kappa}=1.35$ (see text). 
We show also two ``generic'' models G0.1 and G2 with Gaussian $W(z)$ (normalized to provide the whole EGB above 1 GeV and then multiplied by the factor $A^{\gamma\kappa}$ described in the text), with peak at $z_0=0.1$ and width $\sigma_z=0.1$ (cyan-dashed), and $z_0=2$ and $\sigma_z=0.5$ (magenta-dashed), respectively.
In the upper inset, we show the EGB benchmark model and {\sl Fermi-LAT} measurement~\citep{Fermi:2014}. \\ \\
}
\label{fig1}
\vspace{-0.3cm}
\end{figure}

In order to better exploit all the available information encoded in the maps, we can combine the cross-correlation from the different $\gamma$-ray energy bins introduced above.
Since the EGB spectrum roughly scales with $E^{-2.4}$ (see inset in Fig.~\ref{fig1}), low energy bins have larger statistics.
On the other hand, the {\sl Fermi-LAT} point spread function significantly improves at high energy~\citep{Ackermann:2012kna}. We therefore expect an information gain by splitting the signal in different energy bins.
A minimum variance combination of the 6 single $E$-bin $C_b^{(\gamma^{i}\kappa)}$ measurements in a given multipole bin $b$ can be defined as:
\be
C_b^{(\gamma\kappa)} = \sum_{i=1}^6 w_i(b) C_b^{(\gamma^{i}\kappa)}\;,\; w_i(b) = N_{b} \sum_{j=1}^6  [\Gamma^{-1}_b]^{ij}\;,
\label{eq:mv}
\ee
where $\Gamma_b^{ij} \equiv {\rm Cov}[C_b^{(\gamma^{i}\kappa)},C_b^{(\gamma^{j}\kappa)}]$ is the $6 \times 6$ sub-matrix for the covariance in the bin $b$, and $N_{b}=\left( \sum_{ij} [\Gamma^{-1}_b]^{ij}  \right)^{-1}$.
Note that, after having checked the stability of our results against the inclusion of the correlation among different multipole bins $\Gamma_{bb'}^{ij}$, we choose not to include them for simplicity.
 We normalize $C_b^{(\gamma^{i}\kappa)}$ by means of the factor $E_i^{2.4}/\Delta E_i$ (with $E_i=\sqrt{E_{{\rm max},i}E_{{\rm min},i}}$ and $\Delta E_i=E_{{\rm max},i}-E_{{\rm min},i}$) to make it approximately flat in energy.

The computation of the full covariance matrix including correlation among different $E$-bins is not straightforward. 
Whereas the correlation terms among different multipole bins within an $E$-bin are provided by {\sffamily PolSpice}, we estimate the off-diagonal correlation between the $E_i$ and $E_j$ bins (with $i\neq j$) using a two-step process.
We first derive a semi-analytic Gaussian approximation (averaged in the multipole bin $b$):
\be
  \tilde \Gamma_b^{ij} =\left\langle \frac{C_{\ell}^{(\gamma^i\kappa)}C_{\ell}^{(\gamma^j\kappa)}+\hat{C}_{\ell}^{(\gamma^i\gamma^j)}\hat{C}_{\ell}^{(\kappa)}}{(2\ell+1)\,f_{\rm sky}}\right\rangle_b,
\label{eq:dcl}
\ee
where $C_\ell^{(\gamma^i\kappa)}$ is the cross-correlation APS, estimated using a benchmark theoretical prediction discussed in the next Section. 
(Note that this term is in any case subdominant in Eq.~(\ref{eq:dcl})). $\hat{C}_{\ell}^{(\kappa)}$ and $\hat{C}_{\ell}^{(\gamma^i)}$ are the auto-correlation APS that we estimate from the corresponding maps using {\sffamily PolSpice} and $\hat{C}_{\ell}^{(\gamma^i\gamma^j)}$ is the cross-correlation APS between two energy bins $i$ and $j$. 
As a sanity test, we checked that the noise-subtracted estimate $C_\ell^{(\gamma^i)}= \hat{C}_{\ell}^{(\gamma^i)} - (C_N^{(\gamma^i)}/W_{\ell}^2)$ (where $C_N$ is the power spectrum of the shot noise and $W_{\ell}$ is the beam function) agrees well with the autocorrelation APS reported by the {\sl Fermi-LAT} Collaboration~\citep{Ackermann:2012uf}. Similarly, our $\hat{C}_{\ell}^{(\kappa)}$ is consistent with theoretical expectations, once corrected for the noise APS provided in the Planck public data release~\citep{Planck2013lensing}.
The factor $f_{\rm sky}$ corrects for the effective available fraction of the sky, but Eq.~(\ref{eq:dcl}) might actually underestimate the impact of masks.
To have a more conservative error estimate we derive a scaling coefficient $M_{i,b}$ from $\Gamma_b^{ii}=M_{i,b}^2 \,\tilde \Gamma_b^{ii}$, where $\Gamma_b^{ii}$ is obtained from {\sffamily PolSpice} and $\tilde \Gamma_b^{ii}$ from Eq.~(\ref{eq:dcl}), and then define the off-diagonal terms of the covariance matrix as $\Gamma_b^{ij}=M_{i,b}\,M_{j,b} \tilde \Gamma_b^{ij}$. The reliability of this scaling is further supported by the fact we are using the same mask for all the $\gamma$-ray maps.

The combined APS $C_b^{(\gamma\kappa)}$ of Eq.~(\ref{eq:mv}) is shown in Fig.~\ref{fig1} for the four cases considered. Error bars are given by $\sqrt{N_{b}}$.
The different analyses are in excellent agreement with each other.
As for the analysis with gamma-rays integrated above 1 GeV, we estimate the significance of the cross-correlation signal in the multipole-bins $40\leq\ell<160$, $160\leq\ell<280$, $280\leq\ell<400$. The significances now amount to: $3.0\sigma$, $0.7\sigma$ and $1.2\sigma$, respectively. A comparison with the results of the previous analysis shows that by adding spectral information increases the significance of the signal in the low-$\ell$ sector, while in the
larger-$\ell$ bins the cross-correlation are still compatible with zero. The results obtained so far
therefore show an evidence of correlation for multipoles below 
$\ell\lesssim 150-160$.

%%%%%%%%%%%%%%%%%%%%%%
\begin{table*}[t!]
\def\arraystretch{1.3}
\centering
\begin{tabular}{|c|c|c|c|c|c|c|c|c|c|}
\hline
\multicolumn{2}{|c|}{Energy} & \multicolumn{2}{|c|}{Multipole} & Statistical & \multicolumn{4}{|c|} {Significance}\\
\multicolumn{2}{|c|}{} & \multicolumn{2}{|c|}{} & test &P15-3FGL&P15-2FGL&P13-3FGL&P13-2FGL\\
\hline
Single $E$-bin & $[1, 300]$ GeV & 
Single $\ell$-bin & $40\leq\ell<160$ & $\langle \ell C_\ell^{\gamma\kappa}\rangle/\delta\langle \ell C_\ell^{\gamma\kappa}\rangle$ & $1.7\sigma$ & $1.8\sigma$ & $1.5\sigma$ & $2.1\sigma$\\
\hline
6 $E$-bins & $[0.7,300]$ GeV & 
Single $\ell$-bin & $40\leq\ell<160$ & $\langle \ell C_\ell^{\gamma\kappa}\rangle/\delta \langle \ell C_\ell^{\gamma\kappa}\rangle$ & $3.0\sigma$ & $3.3\sigma$ & $2.8\sigma$ & $3.2\sigma$ \\
\hline
6 $E$-bins & $[0.7,300]$ GeV & 
6 $\ell$-bins, $\Delta\ell = 60$& $40\leq\ell<400$ & Model fitting & $3.0\sigma$ & $3.2\sigma$ & $2.7\sigma$ & $3.0\sigma$ \\
\hline
\end{tabular}
\caption{Summary of statistical significances for the three adopted methods. All analyses are performed on $\ell C_\ell^{\gamma\kappa}$, to make the observable approximately flat in multipoles. The errors $\delta\langle \ell C_\ell^{\gamma\kappa}\rangle$ are obtained from the covariance matrix of {\sffamily PolSpice}. In the first row, the symbol $\langle\cdot\rangle$ denotes mean in the multipole bin. In the second row, the APS (and corresponding errors) at different energies $E_i$ are obtained as discussed in connection to Eq.~(\ref{eq:mv}) and are whitened through multiplication by $E_i^{2.4}/\Delta E_i$ (with the symbol $\langle\cdot\rangle$ denoting the average in a multipole bin and among energy bins).
The third row reports model fitting: the significance is obtained from a
$\chi^2$  difference between null signal and  best-fit model. P15 (P13) stands for the analysis using the Planck 2015 (2013) map.\\}
\label{tab:sigmas}
\end{table*}
%%%%%%%%%%%%%%%%%%%%%%

As a cross-check for the stability of $\gamma$-ray data, we repeat the analysis considering the data from the first 150 weeks and subsequent 150 weeks separately. The obtained APS are compatible and, once combined together, very closely resemble the APS of the full period presented above.

The subtraction of the galactic foreground in the $\gamma$-ray maps has a significant systematic uncertainty related to the modeling of the galactic diffuse emission, which can affect anisotropies on large scales~\citep{Ackermann:2012uf}. The foreground residuals in the lensing map are instead thought to be more under control since they do not show up in the autocorrelation studies~\citep{Planck2015lensing}.
Assuming the lensing map to be free from galactic contaminations, the presence of a gamma-ray galactic foreground in the maps would not provide a cross-correlation signal. Rather it would only act as a noise term.
To test this, we performed the same analysis discussed above but employing $\gamma$-ray maps where the foreground was not subtracted. We found the same central values for the cross-correlation APS points, but with larger errors (and so lower statistical significance), consistent with the fact that the galactic foregrounds contribute to the error budget but not to the signal.
This suggests that possible contaminations of the APS from a galactic foreground bispectrum are small.

In the next Section we will show that the derived APS can be explained in terms of gamma-rays emission from astrophysical sources emitting mostly at intermediate redshifts.

%%%%%%%%%%%%%%%%%%%%%%%%%%%%%%%%%%%%%%%%%%%%%%%%%%%%%%%%%%%%%%%%%%%%%%%%%%%%%
\section{Interpretation}
\label{sec:res}
We now move on to discuss the agreement between theoretical models and the measurements reported in Fig.~\ref{fig1}.

In the Limber approximation \citep{limber}, the theoretical two-point cross-correlation APS can be computed as:
\be
 C_\ell^{(\gamma\kappa)} =  \int \frac{d\chi}{\chi^2} W_\gamma(\chi)\,W_\kappa(\chi)\,
 P_{\gamma\kappa}(k=\ell/\chi,\chi)\;.
  \label{eq:Cl}
\ee
where $\chi(z)$ denotes the radial comoving distance, $W_\kappa$ and $W_\gamma$ are the window functions for lensing and $\gamma$ rays, and $P_{\gamma\kappa}$ is the three-dimensional (3D) power-spectrum (PS) of the cross-correlation. For the latter we follow the halo model approach (see, e.g., \citep{Cooray:2002dia} for a review), where $P$ can be split in the one-halo $P_{\rm 1h}$ and two-halo $P_{\rm 2h}$ components as $P=P_{\rm 1h}+P_{\rm 2h}$ (see \citep{Fornengo:2013rga} for their expressions).

The CMB lensing window function is given by~\citep{Bartelmann:2010fz}:
\be
W_\kappa(\chi)=\frac{3}{2}\ho^2\om[1+z(\chi)]\chi\,\frac{\chi_*-\chi}{\chi_*}\;,
\label{eq:wcmb}
\ee
where $\ho$ is the Hubble constant, $\om$ is the matter-density parameter, and $\chi_*$ is the comoving distance to the last-scattering surface.

The window function for a $\gamma$-ray emitter $i$ is (see, e.g., \citep{Camera:2012cj}):
\be
W_{\gamma_i}(E,z) =\frac{\int_{\lum_{\rm min}(z)}^{\lum_{\rm max}(z)} d\lum\,\Phi_{\gamma_i}\,\lum}{4\,\pi\,(1+z)} \, \exp[{-\tau[E(1+z),z]}]\;
\ee
where $\lum$ is the $\gamma$-ray luminosity per unit energy range, $\Phi_\gamma(\lum,z)$ is $\gamma$-ray luminosity-function (GLF), and $\tau$ is the optical depth for absorption~\citep{Stecker:2006eh}.

We consider four different extragalactic $\gamma$-ray populations: star forming galaxies (SFG), misaligned AGN (mAGN), and two subclasses of blazars, BL Lacertae (BL Lac) and flat spectrum radio quasars (FSRQs).
The GLFs of the last three source classes are taken from the best-fit models of, respectively, \citep{DiMauro:2013xta}, \citep{Ajello:2013lka}, and \citep{Ajello:2012}. 
In the case of SFG we consider the infrared luminosity function from \citep{Gruppioni:2013jna} (adding up spiral, starburst, and SF-AGN populations of their Table 8), and linking $\gamma$ and infrared luminosities by means of the relation derived in \citep{Fermi:2012eba}.
The energy spectrum is assumed to be a power-law with spectral indexes $-2.7$ (SFG), $-2.37$ (mAGN), $-2.1$ (BL Lac), and $-2.4$ (FSRQ).
The model fairly reproduces {\sl Fermi-LAT} measurements for both the EGB (see the upper-right inset of Fig.~\ref{fig1}) and the $\gamma$-ray autocorrelation APS.
For the latter, we found a flat APS (given by the 1-halo term and dominated by BL Lac contribution) with $C_\ell=1.5\times 10^{-17} {\rm cm^{-4}\,s^{-2}\,sr^{-1}}$ for $E>1$ GeV.

The cross-correlation power spectrum at the intermediate scales considered here is mostly set by the linear part of the clustering, $P\simeq P_{\rm 2h}$, which is similar in the various cases (i.e., it is related to the linear total matter PS $P_{\rm lin}$) except for the specific bias term, with negligible contribution from $P_{\rm 1h}$. In other words, we approximately have $C_\ell^{(\gamma_i)\kappa}=  \int d\chi\,\chi^{-2}\, W_{\gamma_i}(\chi)\, W_\kappa(\chi)\, b^{\gamma_i}_{\rm eff}(z) P_{\rm lin}(\ell/\chi,\chi)$, where the ``effective'' bias of a $\gamma$-ray population is: $b^{\gamma_i}_{\rm eff}(z)=\int d\lum\,b^{\gamma_i}(\lum,z)\,\Phi_{\gamma_i}\,\lum/(\int d\lum\,\Phi_{\gamma_i}\,\lum)$ with $b^{\gamma_i}(\lum,z)$ being the bias between the $\gamma$-ray source $i$ and matter, as a function of luminosity and redshift.
To estimate the latter, we use the halo bias $b_h$~\citep{Sheth:1999mn} (setting $b^{\gamma_i}(\lum,z)=b_h(M^{\gamma_i}(\lum,z))$ and the relation $M^{\gamma_i}(\lum,z)$ (setting the mass of the halo hosting astrophysical objects $i$ with a certain luminosity $\mathcal{L}$), as described in \citep{Camera:2014rja}.
Comparing the bias of blazars obtained in our analysis with the bias derived in \citep{Allevato:2014}, we found the latter to be somewhat larger than our estimates. We obtain a mean mass hosting the object of $M=5\times 10^{12}M_\odot$ (BL LAC) and $M=1.5\times 10^{13}M_\odot$ (FSRQ) contrary to $M=3\times 10^{13}M_\odot$ of \citep{Allevato:2014}. However, our measurement probes the unresolved (individually fainter) component which reside in less massive halos than the brightest blazar sub-sample considered in \citep{Allevato:2014}. Thus the two results are not in contradiction.

The cross-correlation APS predicted in the models of the four $\gamma$-ray emitters described above and their collective contribution are shown in Fig.~\ref{fig1}.

With the theoretical model at hand we can fit its overall amplitude $A^{\gamma\kappa}$ by minimizing the $\chi^2$, which is computed by means of the full covariance matrix introduced above.
The statistical significance of the model is derived computing the $\Delta \chi^2$ between null signal and best-fit model. We obtain $A^{\gamma\kappa}=1.35\pm0.45$ with 3.0$\sigma$ significance
which shows a statistically significant preference for a signal with the correct features expected from the extragalactic gamma-ray emission.

The window functions of the considered $\gamma$-ray populations are all peaked at $z\sim0.5-1$.
To explore in a more general way the kind of $\gamma$-ray model preferred by the data, we compute in Fig.~\ref{fig1} the signals from two Gaussian window functions $W(z)\propto \exp[-(z-z_0)/\sigma_z^2]$, one peaked at low redshift (model G0.1 with $z_0=\sigma_z=0.1$), and one peaked at high redshift (model G2 with $z_0=2$ and $\sigma_z=0.5$), both normalized to match the {\sl Fermi-LAT} EGB measurement above 1 GeV (and bias modelled as for mAGNs).
We found $A^{\gamma\kappa}_{G0.1}=2.99\pm0.96$ ($3.1\sigma$) and $A^{\gamma\kappa}_{G2}=0.85\pm0.29$ ($2.9\sigma$).
For $W(z)$ peaked at $z\gg1$ the relative contribution of small (more distant) objects with respect to larger objects increases, while no power is detected at small scales (above $\ell\sim150$). This slightly reduces the statistical significance (although with the current data accuracy we cannot exclude this possibility).
On the contrary, $W(z)$ peaked at low $z$ would provide the right bump at low $\ell$, increasing the statistical significance. 
However, the large value of the overall amplitude translates into $\langle b_{\rm eff}\rangle\sim3$, which is typically way too large for a low-$z$ population (see e.g., \citep{Cooray:2002dia}).
Note also that, since the window function of the CMB lensing peaks at moderately high redshift, as mentioned in the Introduction, its overlapping is more effective with high-$z$ $\gamma$-populations rather than low-$z$ emitters. Therefore, in the latter case, the required $\langle b_{\rm eff}\rangle$ becomes slightly larger.

Above arguments seem to suggest that, in order to reproduce the observed cross-correlation, the bulk of $\gamma$-ray contribution to the EGB have to reside at intermediate redshift.

\begin{figure}[t]
\vspace{-3cm}
\includegraphics[width=0.45\textwidth]{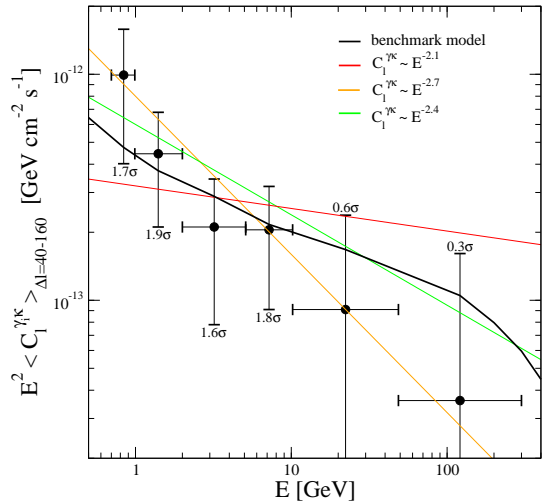}
\caption{Energy dependence of the cross-correlation APS. The reported points are for $E^{2}\langle C_\ell^{\gamma_i\kappa}\rangle_{\Delta\ell=40-160}$, which is the energy-differential APS  of each energy bin $i$ averaged in the multipole bin $40<\ell<160$ and multiplied by $E^2$. Errors are from the diagonal of the covariance matrix and we report the statistical significance of each point.
The benchmark theoretical model is shown in black and is multiplied by the energy averaged amplitude $A^{\gamma\kappa}=1.35$. \\ \\}
\label{fig2}
\vspace{-0.3cm}
\end{figure}

Fig.~\ref{fig2} shows the measured cross-correlation APS for different energy bins and averaged in the multipole bin $40<\ell<160$. 
The spectrum is consistent with the benchmark model and similar to the {\sl Fermi-LAT} EGB spectrum (having spectral index close to $-2.4$), although possibly slightly softer.

%%%%%%%%%%%%%%%%%%%%%%%%%%%%%%%%%%%%%%%%%%%%%%%%%%%%%%%%%%%%%%%%%%%%%%%%%%%%
%%%%%%%%%%%%%%%%%%%%%%%%%%%%%%%%%%%%%%%%%%%%%%%%%%%%%%%%%%%%%%%%%%%%%%%%%%%%
\section{Discussion and Conclusions}
\label{sec:discussion}

We reported the first indication of a cross-correlation between the unresolved $\gamma$-ray sky and CMB lensing. The analysis also points towards a direct evidence that the IGRB is of extragalactic origin.
The analysis has been based on the $\gamma$-ray data of the first 68 month of operation of the {\sl Fermi-LAT} and on the 2013 public release by the \Planck\ Collaboration of the CMB lensing potential map. Current models of AGN and SFG can fit well the amplitude, angular dependence and energy spectrum of the observed APS. 
The size of the signal appears to be robust against variations of the analysis assumptions.
Data exhibit a preference for a signal with the correct features expected from the extragalactic gamma-ray emission with a $3.0\sigma$ significance.

The forthcoming {\sl Fermi-LAT} Pass-8 reprocessed events will allow for a more refined assessment of the signal.
Moreover, the technique presented in this work can be also applied for 
cross-correlating the $\gamma$-ray sky with probes of the large scale structure of the
Universe at different redshifts (such as galaxy catalogues and weak-lensing surveys).
Such a tomographic analysis of the EGB will provide invaluable information about its composition.

Contaminations from foreground, either real (e.g., a dust or point sources bi-spectrum) or spurious, cannot, at present, be totally excluded, but they are significantly disfavoured.
A model of $\gamma$-ray populations built to explain the EGB, and not tuned to the measurement presented here, matches well the data both in features and normalization.
More generically, a population of extragalactic $\gamma$-ray emitters following matter clustering at large scales with GLF peaked at intermediate $z$ and with $\langle b_{\rm eff}\rangle\sim 2-3$ agrees well with the data, once the associated EGB is normalized to fit the {\sl Fermi-LAT} measurement of the IGRB. On the contrary, if, for example, the contribution to the IGRB is reduced to 50\%, the required bias would become $\langle b_{\rm eff}\rangle\sim 4-6$, which is likely unrealistically large. 
This implies that the presented results can be considered as a first direct proof that the majority of the IGRB is emitted by extragalactic structures. 

%%%%%%%%%%%%%%%%%%%%%%%%%%%%%%%%%%%%%%%%%%%%%%%%%%%%%%%%%%%%%%%%%%%%%%%%%%%%%

\bigskip
\acknowledgments
We thank D.~Maurin for a question which was the actual kick-off of the project, and A.~Cuoco, M.~Fornasa and H.~Zechlin for insightful discussions. This work is supported by the research grant {\sl Theoretical Astroparticle Physics} number 2012CPPYP7 under the program PRIN 2012 funded by the Ministero dell'Istruzione, Universit\`a e della Ricerca (MIUR), by the research grants {\sl TAsP (Theoretical Astroparticle Physics)} and Fermi funded by the Istituto Nazionale di Fisica Nucleare (INFN), and by the  {\sl Strategic Research Grant: Origin and Detection of Galactic and Extragalactic Cosmic Rays} funded by Torino University and Compagnia di San Paolo.

\end{document}